\documentclass[twocolumn,showpacs,amsmath,nofootinbib,amssymb,epsfig]{revtex4}

\usepackage{graphicx}
\usepackage{dcolumn}
\usepackage{bm}

\begin{document}
\title{Einstein Static Universe in  Braneworld Scenario}

\author{K. Atazadeh}
\email{atazadeh@azaruniv.ac.ir}\affiliation{Department of Physics, Azarbaijan Shahid Madani University , Tabriz, 53714-161 Iran\\Research Institute for Astronomy and Astrophysics of Maragha (RIAAM),
Maragha 55134-441, Iran}
\author{Y. Heydarzade}
\email{heydarzade@azaruniv.ac.ir}\affiliation{Department of Physics, Azarbaijan Shahid Madani University , Tabriz, 53714-161 Iran\\Research Institute for Astronomy and Astrophysics of Maragha (RIAAM),
Maragha 55134-441, Iran}
\author{F. Darabi}
\email{f.darabi@azaruniv.ac.ir}\affiliation{Department of Physics, Azarbaijan Shahid Madani University , Tabriz, 53714-161 Iran\\Research Institute for Astronomy and Astrophysics of Maragha (RIAAM),
Maragha 55134-441, Iran}

\date{\today}

\begin{abstract}
The stability of Einstein static universe against homogeneous
scalar perturbations in the context of braneworld scenario is investigated.
The stability regions are obtained  in terms of the constant geometric linear equation of state parameter $\omega_{extr}=p_{extr}/\rho_{extr}$ and are studied for each evolutionary era of the universe. The results are discussed for the case of closed, open or flat universe in each era under the obtained restricting conditions. We also briefly investigate the stability against
vector and tensor perturbations. Contrary to the classical general relativity, it is found that a stable Einstein static universe may exist in a braneworld theory of gravity against scalar, vector and tensor perturbations for some
suitable values and ranges of the cosmological parameters.
\end{abstract}
\maketitle

\section{Introduction}
First attempts for finding a static solution of the field equations of general
relativity to describe a homogenous and isotropic universe was done by Einstein. Since the Einstein filed equations have no static solution, the so called cosmological constant was introduced by Einstein  to make the solutions  static \cite{Einstein}. Thereafter,
 it was shown by Eddington that the Einstein static universe is unstable against the spatially homogeneous and isotropic perturbations \cite{Eddington}. Later work done by Harrison showed that in a radiation-filled Einstein static universe, all the physical inhomogeneous modes are oscillatory \cite{Harrison}.
Also, for the Einstein static universe, Gibbons showed that the entropy is
maximized for an equation of state with the sound speed satisfying $c_{s}\equiv
dp/d\rho>1/\sqrt5$ \cite{Gibbons}. These  results have been further investigated by Barrow {\it et.al} \cite{Barrow2003}, where
it was shown that Einstein static universe
is always neutrally stable against small inhomogeneous vector and tensor perturbations and also neutrally stable against inhomogeneous adiabatic scalar density perturbations with the sound speed  $c_{s}>1/\sqrt5$ . Recently, it was
shown that the Einstein static universe is unstable against Bianchi type-IX spatially homogeneous perturbations in the presence of tilted and non-tilted perfect fluid with $\rho+3P>0$ and for some kinds of matter field sources \cite{Barrow2009, Barrow2012}.

A renewed motivation for studying the Einstein static universe comes from the
emergent universe scenario \cite{Ellis}.
This cosmological model is
 a past-eternal inflationary model in
which the horizon problem is solved before the beginning of inflation and
 the  big-bang singularity is removed. Also, in this cosmological model no exotic physics is involved and the quantum gravity regime can even be avoided. The inflationary universe emerges from a
small static state containing the seeds for the development of the microscopic universe. However, this cosmological model suffers from a fine-tuning problem which can be ameliorated by modifications to the cosmological equations
of general relativity.
For this reason, analogous static solutions have been explored in the context of different modified theories of gravity. For instance, the Einstein static universe has been analyzed in $f(R)$ gravity \cite{f(R)a, f(R)b, f(R)c}, $f(T)$ gravity \cite{f(T)}, Einstein-Cartan theory \cite{Bohmer} and nonconstant pressure models \cite{Pressure}. Also, this model is studied in the    Horava-Lifshitz  gravity \cite{Horava}, IR modified Horava gravity \cite{HoravaIR} and loop quantum cosmology \cite{Loop}. In addition, this model has been studied in braneworld models inspired by string/M theory  in which gravity is a truly higher-dimensional theory and becomes effectively 4-dimensional at lower energies. In these models, the standard gauge interactions are confined to the four-dimensional space time (the braneworld generated by a 3-brane) embedded in higher dimensional bulk, while the gravitational field probes the extra dimensions \cite{Arkani, Randall, Dvali}( see also \cite{Maartens,
Langlois} for a review on brane gravity). As an instance of studying the Einstein static universe in the framework of braneworld scenarios, the authors of \cite{Gergely} explored braneworld generalizations of the Einstein static universe. It was shown that a static Friedmann brane in a 5-dimensional bulk (Randall-Sundrum type model) can have a very different relation between the density, pressure, curvature and cosmological constant than the case of the general relativistic Einstein static universe. In particular, static Friedmann branes with zero cosmological constant and 3-curvature, but satisfying $\rho > 0$ and $\rho + 3p > 0$, are shown to be possible.
Also, the stability of an Einstein static universe in the DGP braneworld scenario is analyzed in \cite{Zhang}. This model was divided into two separate branches denoted by $\epsilon = \pm 1$ . The $\epsilon= +1$ branch can explain the present accelerated cosmic expansion without the introduction of dark energy, while for the $\epsilon= -1$ branch, dark energy is needed in order to yield an accelerated expansion. Assuming the existence of a perfect fluid with a constant equation of state $\omega$, the authors find that: i) for the $\epsilon= 1$ branch, there is no a stable Einstein static solution,
and ii) for the  $\epsilon = -1$ branch, the Einstein static universe exists and it is stable for  $ -1 < \omega < -1/3$. Thus, the universe can stay at this stable state past-eternally and may undergo a series of infinite, nonsingular oscillations. Therefore, the big bang singularity problem in the standard cosmological model can be resolved. An oscillating universe in the DGP braneworld scenario is also studied in \cite{Zhangb}.
By assuming that the energy component is a pressureless matter, radiation or vacuum energy, respectively, the authors find that in the matter or vacuum energy dominated case, the scale factor has a minimum value $a_0$. In the matter dominated case, the big bang singularity can be avoided in some special
circumstances, and there may exist an oscillating universe or a bouncing one. In the vacuum energy dominated case, there exists a stable Einstein
static state to avoid the big bang singularity. However, in certain circumstances in the matter or vacuum energy dominated case, a new kind of singularity may occur at $a_0$ as a result of the discontinuity of the scale factor. In the radiation dominated case, the universe may originate from the big bang singularity, but a bouncing universe which avoids this singularity is also possible. Moreover, the authors of \cite{Clarkson} discussed the Einstein static brane in a Schwarzschild-anti-de Sitter bulk spacetime under tensor perturbations.

In the present work, we investigate the stability of Einstein static universe against homogeneous scalar, vector and tensor perturbations in the context of braneworld scenario where the effective field equations are induced on the brane. This work is based on the  model studied in \cite{Maia} where a geometrical interpretation for dark energy as warp in the universe given by the extrinsic curvature was proposed. The induced field equations on the brane are studied with respect to the perturbation in the cosmic scale factor $a(t)$, where the confined energy density $\rho(t)$ depends only on time. We consider the evolution of field equations up to the linear perturbations and neglect all higher order terms. The stability regions are obtained in terms of constant geometric linear equation of state parameter $\omega_{extr}=p_{extr}/\rho_{extr}$ for each evolutionary era of the universe. We discuss about the results  for the case of closed, open or flat universe in each era under the obtained restricting conditions. Throughout this paper, we use the units for which
$8\pi G=1$.
\section{The model}
Based on the  model proposed in \cite{Maia}, the induced Einstein equation,
 modified by the presence of the extrinsic curvature, on $4D$ brane is as
follows
\begin{equation}\label{21}
G_{\mu\nu}=T_{\mu\nu}-\Lambda g_{\mu\nu}+Q_{\mu\nu},
\end{equation}
where $T_{\mu\nu}$ and  $\Lambda$ are  the confined source and the effective cosmological constant of the four dimensional brane, respectively.
Also, $Q_{\mu\nu}$ is a pure geometrical quantity
as
\begin{eqnarray}\label{22}
Q_{\mu\nu}&=&{K^{\rho}}_{\mu}K_{\rho\nu}-g^{\lambda \rho}K_{\lambda \rho}K_{\mu\nu}
-\\\nonumber&&\frac{1}{2}
(K^{\rho\lambda}K_{\rho\lambda}-g^{\lambda\nu}g^{\alpha\beta}K_{\lambda\nu}K_{\alpha\beta})
g_{\mu\nu},
\end{eqnarray}
 where the $K_{\mu\nu}$ and $g_{\mu\nu}$ are the extrinsic curvature and
the $4D$ brane
metric, respectively. For the purpose of  embedding of the $FRW$ brane in a five dimensional bulk space, one should consider the metric
\begin{equation}\label{23}
 ds^{2}=-dt^{2}+a(t)^{2}\left(\frac{dr^2}{1-kr^2}+r^{2}d\Omega^2\right),
 \end{equation}
where $a(t)$ is the cosmic scale factor and $k=+1, -1$ or $0$ corresponds
to the closed, open or flat universes, respectively.
Also, the confined source to the brane $T_{\mu\nu}$ can be considered as a perfect fluid given in co-moving coordinates
by\begin{equation}\label{24}
 T_{\mu\nu}=(\rho + p)u_{\mu}u_{\nu}+ pg_{\mu\nu},
 \end{equation}
where $u_{\alpha}=\delta^{0}_{\alpha}$, and $\rho$, $p$ are energy density
and isotropic pressure, respectively.

The calculations carried out in \cite{Maia} yield the extrinsic curvature profiles as
 \begin{eqnarray}\label{25}
 &&K_{00}=-\frac{1}{\dot a}\frac{d}{dt}\left(\frac{b}{a}\right),\nonumber\\
 &&K_{ij}=\frac{b}{a^2}g_{ij},\ i,j=1,2,3.
 \end{eqnarray}
where dot means derivative with respect to the cosmic time $t$ and $b=b(t)$ is an arbitrary function. By defining the parameters $h:=\frac{\dot b}{b}$ and $H:=\frac{\dot a}{a}$ the components of $Q_{\mu\nu}$ represented by (\ref{22}) take the form of
\begin{eqnarray}\label{26}
&& Q_{00}=\frac{3b^2}{a^4},\nonumber\\
&&Q_{ij}=-\frac{b^2}{a^4}\left(\frac{2h}{H}-1\right)g_{ij}.
\end{eqnarray}
Similar to the confined source $T_{\mu\nu}$, the geometric energy-momentum tensor $Q_{\mu\nu}$ can be identified as
\begin{equation}\label{27}
 Q_{\mu\nu}=(\rho_{extr} + p_{extr})u_{\mu}u_{\nu}+ p_{extr}g_{\mu\nu},
\end{equation}
where $\rho_{extr}$ and  $p_{extr}$ denote the ``geometric energy density"
and ``geometric pressure", respectively (the suffix $extr$ stands for "extrinsic").
Then, using the equations  (\ref{26}) and (\ref{27}) we obtain
 \begin{eqnarray}\label{28}
 &&\rho_{extr}=\frac{3b^2}{a^4},\nonumber\\
 &&p_{extr}=-\frac{b^2}{a^4}\left(\frac{2h}{H}-1\right).
 \end{eqnarray}
Moreover, the geometric fluid can be implemented by the equation of state $p_{extr}=\omega_{extr}\rho_{extr}$ where  $\omega_{extr}$ is the geometric
equation of state parameter and generally can be a function of time \cite{Maia}. Using
 equations (\ref{28}) and the equation of state of the geometric fluid,
we obtain the following equation for $b(t)$
 \begin{equation}\label{29}
\frac{\dot b}{b}=\frac{1}{2}\left(1-3\omega_{extr}\right)\frac{\dot a}{a},
\end{equation}
which cannot be readily solved because $\omega_{extr}$ is not known.
However, in the study of Einstein static universe, a simple and useful case may be considered as $\omega_{extr}=\omega_{0extr}={constant}$, which leads to a general solution of (\ref{29}) as
\begin{equation}\label{210}
b=b_{0}\left(\frac{a}{a_0}\right)^{\frac{1}{2}(1-3\omega_{0extr})},
\end{equation}
where $a_0=constant$ is the scale factor of Einstein static universe and $b_0$ is an integration constant related to the curvature warp of this
universe. Substituting  equation (\ref{210}) into  equations (\ref{26})
gives the geometric fluid component in terms of $b_0$, $a_0$ and $a(t)$ as
\begin{eqnarray}\label{211}
&&Q_{00}(t)=\frac{3b_{0}^{2}}{a_{0}^{1-3\omega_{extr}}}a^{-3(1+\omega_{extr})},\nonumber\\
&&Q_{ij}(t)=3\omega_{extr}\frac{b_{0}^{2}}{a_{0}^{1-3\omega_{extr}}}a^{-3(1+\omega_{extr})}g_{ij},
\end{eqnarray}
and consequently using equations (\ref{28})
we get
\begin{eqnarray}\label{212}
&&\rho_{extr}(t)=\frac{3b_{0}^{2}}{a_{0}^{1-\omega_{extr}}}a^{-3(1+\omega_{extr})},\nonumber\\
&&p_{extr}(t)=3\omega_{extr}\frac{b_{0}^{2}}{a_{0}^{1-3\omega_{extr}}}a^{-3(1+\omega_{extr})}.
\end{eqnarray}
For the Einstein static universe, $a=a_{0}=constant$, the  geometric
fluid components are
\begin{eqnarray}\label{213}
&&Q_{00}(a_0)=\frac{3b_{0}^2}{a_{0}^4},\nonumber\\
&&Q_{ij}(a_0)=3\omega_{extr}\frac{b_{0}^2}{a_{0}^4}g_{ij}.
\end{eqnarray}
Consequently, using equations (\ref{212}), the  geometric energy density and isotropic pressure take the form
of
\begin{eqnarray}\label{214}
&&\rho_{0extr}=\rho_{extr}(a_0)=\frac{3b_{0}^2}{a_{0}^4},\nonumber\\
&&p_{0extr} =p_{extr}(a_{0})=\frac{3\omega_{extr}b_{0}^2}{a_{0}^4}.
\end{eqnarray}
Using equations (\ref{24}) and (\ref{211}), the induced Einstein equation on the brane (\ref{21}) give us the following equation for the confined energy
density
\begin{equation}\label{215}
\rho(t)=3\left(\frac{\dot a}{a}\right)^{2}+\frac{3k}{a^2}-\frac{3b_{0}^2}{a_{0}^{1-3\omega_{extr}}}a^{-3(1+\omega_{extr})}-\Lambda,
\end{equation}
which takes the following value for the Einstein static universe
\begin{equation}\label{216}
\rho_{0}=\rho(a_0)=\frac{3k}{a_{0}^2}-\frac{3b_{0}^2}{a_{0}^4}-\Lambda.
\end{equation}
Similarly, the confined isotropic pressure component can be obtained from equations (\ref{21}), (\ref{24}) and (\ref{211}) as
\begin{equation}\label{217}
p(t)=-2\frac{\ddot a}{a}-\left(\frac{\dot a}{a}\right)^{2}-\frac{k}{a^2}-\frac{3b_{0}^{2}\omega_{extr}}{a_{0}^{1-3\omega_{extr}}}a^{-3\left(1+\omega_{extr}\right)}+\Lambda,
\end{equation}
which leads to
\begin{equation}\label{218}
p_{0}=p(a_0)=-\frac{k}{a_{0}^2}-\frac{3b_{0}^{2}\omega_{extr}}{a_{0}^4}+\Lambda,
\end{equation}
for the Einstein static universe. We have also the following equation for the Einstein static universe
\begin{equation}\label{218'}
\dot{H}=-\frac{1}{2}[\rho_0(1+\omega_0)+\rho_{0extr}(1+\omega_{0extr})]+\frac{k}{a_{0}^2}=0.
\end{equation}

\section{Scalar perturbations}

In what follows, we first consider the linear homogeneous scalar perturbations around the Einstein static universe, given in equations (\ref{216}) and (\ref{218}), and then explore their stability against these perturbations. The perturbations in the cosmic scale factor $a(t)$ and the confined energy density $\rho(t)$ depend only on time and can be represented by
\begin{eqnarray}\label{319}
&&a(t)\rightarrow a_{0}(1+\delta a(t)),\nonumber\\
&&\rho(t)\rightarrow \rho_{0}(1+\delta \rho(t)).
\end{eqnarray}
Substituting these equations in equation (\ref{215}), subtracting $\rho_0$
and linearizing the result, gives the following equation
\begin{equation}\label{320}
\rho_{0}\delta \rho(t)=\left(-\frac{6k}{a_{0}^2}+\frac{9b_{0}^{2}(1+\omega_{extr})}{a_{0}^{4}}\right)\delta a(t).
\end{equation}

Similarly, one can consider a linear
equation of state $p(t)=\omega\rho(t)$ for confined source. Applying the
above mentioned method (for obtaining equation (\ref{320})) on equations (\ref{217}) and (\ref{218}) results in
\begin{equation}\label{321}
\omega \rho_{0}\delta \rho=-2\delta \ddot a +\left(\frac{2k}{a_{0}^2}+
\frac{9b_{0}^{2}\omega_{extr}(1+\omega_{extr})}{a_{0}^{4}}\right)\delta
a.
\end{equation}
Substituting equation (\ref{320}) in (\ref{321}) gives the equation

\begin{eqnarray}\label{322}
&&\delta\ddot a+ \frac{1}{a_{0}^2}\left[-k(1+3\omega)+\right.\\\nonumber&&\left.\frac{9b_{0}^2}{2a_{0}^2}\left(\omega-\omega_{extr}+\omega\omega_{extr}-\omega_{extr}^{2}\right)\right]\delta a=0.
\end{eqnarray}
This equation has the solution
\begin{equation}\label{323}
\delta a=C_{1}e^{iAt}+C_{2}e^{-iAt},
\end{equation}
where $C_1$ and $C_2$ are integration constants and $A$ is given by
\begin{equation}\label{324}
A^{2}=-k(1+3\omega)+\frac{9b_{0}^2}{2a_{0}^2}\left(\omega-\omega_{extr}+\omega\omega_{extr}-\omega_{extr}^{2}\right).
\end{equation}
Then, for having oscillating perturbation modes representing the existence of a stable Einstein static universe, the following condition should be satisfied
\begin{equation}\label{325}
-k(1+3\omega)+\frac{9b_{0}^2}{2a_{0}^2}\left( \omega-\omega_{extr}+\omega\omega_{extr}-\omega_{extr}^{2}\right)>0,
\end{equation}
which can be rewritten as
\begin{equation}\label{326}
\omega_{extr}^{2}+\omega_{extr}(1-\omega)-\omega+\frac{2k(1+3\omega)a_{0}^{2}}{9b_{0}^2}<0,
\end{equation}
leading to the following acceptable range
\begin{equation}\label{327}
\omega_{extr}^{(1)}<\omega_{extr}<\omega_{extr}^{(2)},
\end{equation}
where
\begin{eqnarray}\label{328}
&&\omega_{extr}^{(1)}=-\frac{1}{2}+\frac{\omega}{2}-\frac{1}{2}\sqrt{
(1+\omega)^{2}-\frac{8k(1+3\omega)a_{0}^2}{9b_{0}^2}},\nonumber\\
&&\omega_{extr}^{(2)}=-\frac{1}{2}+\frac{\omega}{2}+\frac{1}{2}\sqrt{
(1+\omega)^{2}-\frac{8k(1+3\omega)a_{0}^2}{9b_{0}^2}}.
\end{eqnarray}
For the values of $\omega_{extr}$ out of the above range, there are no oscillatory modes and consequently there is no a stable Einstein static universe. The acceptable range  for $\omega_{extr}$ can be studied in each era of universe's evolution corresponding to the case of closed, open and flat universe. In next sections, we explore the specific evolutionary states and obtain some
additional  restricting conditions for having a stable Einstein static universe
during each era.
 \subsection{Vacuum energy dominated era}
For the vacuum energy dominated era with equation of state parameter $\omega=-1$, one can obtain
\begin{eqnarray}\label{329}
&&\omega_{extr}^{(1)}=-1-\frac{2}{3}\sqrt{\frac{ka_{0}^2}{b_{0}^2}},\nonumber\\
&&\omega_{extr}^{(2)}=-1+\frac{2}{3}\sqrt{\frac{ka_{0}^2}{b_{0}^2}},
\end{eqnarray}
in which the corresponding acceptable range for $\omega_{extr}$ is
\begin{equation}\label{330}
-1-\frac{2}{3}\sqrt{\frac{ka_{0}^2}{b_{0}^2}}<\omega_{extr}<-1+\frac{2}{3}\sqrt{\frac{ka_{0}^2}{b_{0}^2}}.
\end{equation}
It turns out that for the case of vacuum energy dominated era, the stable
Einstein static universe can not be open, $k=-1$. Therefore, for this era, an stable universe should be flat or closed. It is also interesting to note
that for the case of flat universe $k=0$, the geometric fluid equation of state parameter is equal to the confined vacuum energy equation of state parameter $\omega_{extr}=\omega=-1$.

 \subsection{Radiation dominated era}
 For the radiation dominated
era, $\omega=\frac{1}{3}$, equations (\ref{328}) takes the form of
\begin{eqnarray}\label{331}
&&\omega_{extr}^{(1)}=-\frac{1}{3}-\frac{2}{3}\sqrt{1-\frac{ka_{0}^2}{b_{0}^2}},\nonumber\\
&&\omega_{extr}^{(2)}=-\frac{1}{3}+\frac{2}{3}\sqrt{1-\frac{ka_{0}^2}{b_{0}^2}},
\end{eqnarray}
which through the equation (\ref{327}) leads to the acceptable range
\begin{equation}\label{332}
-\frac{1}{3}-\frac{2}{3}\sqrt{1-\frac{ka_{0}^2}{b_{0}^2}}<\omega_{extr}<
-\frac{1}{3}+\frac{2}{3}\sqrt{1-\frac{ka_{0}^2}{b_{0}^2}}.
\end{equation}
It is also seen that we should have
\begin{equation}\label{333}
1-\frac{ka_{0}^2}{b_{0}^2}>0,
\end{equation}
which reveals that the stable Einstein static universe can be closed, open or flat universe during this evolutionary era. This equation also restricts the scale factor of Einstein static universe $a_0$ and the curvature warp of this universe $b_0$.
\subsection{Matter dominated era}
For the case of matter dominated era corresponding to $\omega=0$, we have \begin{eqnarray}\label{334}
&&\omega_{extr}^{(1)}=-\frac{1}{2}-\frac{1}{2}\sqrt{1-\frac{8ka_{0}^2}{9b_{0}^2}},\nonumber\\
&&\omega_{extr}^{(2)}=-\frac{1}{2}+\frac{1}{2}\sqrt{1-\frac{8ka_{0}^2}{9b_{0}^2}},
\end{eqnarray}
leading to the acceptable range as
\begin{equation}\label{335}
-\frac{1}{2}-\frac{1}{2}\sqrt{1-\frac{8ka_{0}^2}{9b_{0}^2}}<\omega_{extr}<\\
-\frac{1}{2}+\frac{1}{2}\sqrt{1-\frac{8ka_{0}^2}{9b_{0}^2}}.
\end{equation}
The restricting condition corresponding to this case is
\begin{equation}\label{336}
1-\frac{8ka_{0}^2}{9b_{0}^2}>0,
\end{equation}
which, similar to the previous case, represents the point that the stable Einstein static universe can be a closed, open or flat universe during this evolutionary era.

\section{Vector and tensor perturbations}

In the cosmological context, the vector perturbations of a perfect fluid are governed by the comoving dimensionless {\it vorticity} defined as ${\varpi}_a=a{\varpi}$, whose modes satisfy the following propagation equation \cite{8}
\begin{equation}\label{337}
\dot{\varpi}_{\kappa}+(1-3c_s^2)H{\varpi}_{\kappa}=0,
\end{equation}
where $c_s^2=dp/d\rho$ is the sound speed and $H$ is the Hubble parameter.
Note that this equation is valid in our treatment of Einstein static universe in the braneworld scenario, because our field equations are reduced on the
brane as effective 3+1 dimensional Friedmann equations whose effective fluid is a combination of matter fluid $\rho$ and $\rho_{extr}$. For the Einstein static universe with $H=0$, equation (\ref{337}) reduces to
\begin{equation}\label{338}
\dot{\varpi}_{\kappa}=0.
\end{equation}
This indicates that initial vector perturbations remain frozen, so
we have neutral stability against vector perturbations for all equations
of state on all scales in the present formulation of braneworld scenario.

Tensor perturbations, namely gravitational-wave perturbations, of a perfect
fluid with density $\rho$ and pressure $p=\omega \rho$ is described by the comoving dimensionless transverse-traceless shear $\Sigma_{ab}=a\sigma_{ab}$, whose modes satisfy \cite{8}
\begin{equation}\label{339}
\ddot\Sigma_{\kappa}+3H\dot\Sigma_{\kappa}+\left[\frac{\kappa^2}{a^2}+\frac{2k}{a^2}-\frac{(1+3\omega)\rho+2\Lambda}{3}\right]\Sigma_{\kappa}=0,
\end{equation}
where $\kappa$ is the comoving index ($D^2\rightarrow -\kappa^2/a^2$, $D^2$ being
the covariant spatial Laplacian). For the Einstein static universe this equation
reduces to
\begin{equation}\label{340}
\ddot\Sigma_{\kappa}+\left(\frac{{\kappa^2}}{2k}+1\right)[\rho_0(1+\omega)+\rho_{0extr}(1+\omega_{extr})]\Sigma_{\kappa}=0,
\end{equation}
where we have used equations (\ref{214}), (\ref{215}), (\ref{217}) and (\ref{218'}).
This equation indicates that the neutral stability for tensor perturbations
is generally available, except for those values of parameters $k$, $\omega$, and $\omega_{extr}$ for which the multiplication factor in front of $\Sigma_{\kappa}$ becomes negative.
\section{Concluding Remarks}

We have studied the stability of Einstein static universe against the
homogeneous scalar perturbations in the context of braneworld scenario.
Indeed, the induced field equations on the brane have been studied against the perturbations in the time dependent cosmic scale factor $a(t)$ and the time dependent confined energy density $\rho(t)$.
We have considered the evolution of field equations up to linear perturbations and neglected all higher order terms.
We have obtained the stability regions in terms of constant geometric linear equation of state parameter $\omega_{extr}=p_{extr}/\rho_{extr}$ and then studied them for each evolutionary era of the universe.
Moreover, we have discussed about the results for the case of closed, open
or flat universe in each era using the obtained restricting conditions.
It is shown that for the case of vacuum energy dominated era, the stable
Einstein static universe can not be open, $k=-1$, while for the case of radiation and matter dominated era, the stable Einstein static universe can be closed, open or flat. We have also investigated the stability against vector and tensor perturbations. It turns out that neutral stability is granted for
vector perturbations but for tensor perturbations the neutral stability may
be lost for some ranges of the cosmological  parameters $k$, $\omega$, and $\omega_{extr}$.
\section*{Acknowledgments}
This work has been supported financially by Research Institute for Astronomy and Astrophysics of
Maragha (RIAAM) under research project No.1/2782-61.


\end{document}